\documentstyle[aps,prl,amssymb,epsfig,epic]{revtex}
\def\f#1#2{\frac{#1}{#2}}
\def\r#1{{\rule[-2.5mm]{0mm}{7mm}#1}}
\def\vc#1{{\bf{#1}}}
\newif\iffigure

\begin{document}

% change \figurefalse to \figuretrue to include figures
\figuretrue
%\figuretrue

\draft
\twocolumn[\hsize\textwidth\columnwidth\hsize\csname %
@twocolumnfalse\endcsname

\title{Spin polaron in a magnetic field}
\author{D.~Veberi\v c$\,^{1,*}$, P.~Prelov\v sek$\,^{1,2}$,
and I.~Sega$\,^1$}
\address{$^1$~Jozef Stefan Institute, SI-1001 Ljubljana, Slovenia\\
$^2$~Faculty of Mathematics and Physics, University of Ljubljana,
SI-1000 Ljubljana, Slovenia}
%\date{\today}
\maketitle

\begin{abstract}
\widetext
\smallskip
The influence of the homogeneous magnetic field on a single mobile
hole in a magnetic insulator, as represented by the two-dimensional
$t$-$J$ model, is investigated by considering the coupling of the
field to the orbital current. The energy of the $J=0$ system is
analyzed via the high-temperature expansion and the small system
diagonalization. The susceptibility is shown to be diamagnetic and
diverging at low temperatures $T$. In contrast, in the
antiferromagnetic $J>0$ case small systems generically reveal a
tendency towards a paramagnetic response in larger fields at low
$T$. By employing at $T=0$ the cumulant expansion we study the ground
state in arbitrary $B$, showing a behavior very sensitive to the
character of the quasiparticle dispersion and the magnetic-field
strength. At low $B$ the perturbation and small-systems results are
consistent with a pronounced diamagnetic susceptibility at $T\to 0$,
but indicate on a suppressed contribution at intermediate $T\sim J$.
\end{abstract}
]

\narrowtext
\section{Introduction}
In a system of correlated electrons the external magnetic field can
induce several interesting effects. Theoretically the least understood
are those phenomena, where the magnetic field couples to the motion of
charge carriers. In recent years it has been realized that the
understanding of the anomalous temperature- and doping-dependence of
the Hall effect \cite{ong} is one of the most challenging questions in
connection with the normal state of cuprates, being representatives of
two-dimensional (2D) doped magnetic insulators. Here even the sign of
the effect is theoretically controversal \cite{brin1,assa}. The
diamagnetic contribution to the d.c.\ susceptibility has been much
less investigated \cite{wals}, although it is closely related to the
Hall conductivity \cite{rojo}. It emerges from the orbital motion of
mobile carriers. For noninteracting electrons the contribution
corresponds to the Landau diamagnetism, which is largely temperature
independent. In analogy to the Hall effect and other anomalous
properties of the normal state in cuprates, one could expect anomalies
also in the diamagnetic contribution. So far, however, both
experimental and theoretical answers are lacking.

Magnetic field dependence of the eigenstates of tightbinding electrons
is nontrivial even in the absence of any electron correlations
\cite{hofs}, in particular when the dependence of the ground state on
the field strength $B$ and electron density is investigated
\cite{hase,nori}. There have been only few analogous studies of
correlated systems. Recently, the ground state of a single hole in the
2D $t$-$J$ model in the presence of magnetic field \cite{bera} has
been studied. The main message is that for finite (but not very small)
$B$ the energy is reduced by an amount proportional to parameter $t$,
and the result in such a doped insulator was interpreted in terms of
the composite nature of quasiparticles (QP) \cite{bera,bera1}.
Another evident observation is, however, the difficulty to extract a
reasonable result from studies of small systems.

The aim of this paper is to elaborate on the problem of a single hole in
a magnetic insulator in the presence of a homogeneous magnetic field.
We study the planar $t$-$J$ model \cite{rice} as a prototype model for
strongly correlated electrons and electronic properties of
cuprates,
\begin{equation}
H=-t\sum_{\langle ij\rangle s}({\rm e}^{i \theta_{ij}}
\tilde{c}^\dagger_{js}\tilde{c}^{\phantom{\dagger}}_{is}+
\text{H.c.})+J\sum_{\langle ij\rangle}\vc{S}_i\!\cdot\!\vc{S}_j ,
\label{model}
\end{equation}
where $\tilde{c}^\dagger_{is},\tilde{c}_{is}$ are fermionic operators,
projecting out states with the double occupancy. We consider the
system in a homogeneous field $B$, perpendicular to the plane, and use
for convenience the Landau gauge, where
\begin{equation}
\theta_{ij}=\f{e}{\hbar}\vc{d}_{ij}\!\cdot\!\vc{A}(\vc{r}_i),\qquad
\vc{A}=B(0,x,0),
\label{theta}
\end{equation}
with $\vc{d}_{ij}=\vc{r}_j-\vc{r}_i$. The relevant parameter for the
strength of $B$ is the dimensionless flux per plaquette $\alpha=2\pi
Ba_0^2 /\phi_0$, where $\phi_0=h/e$ is the unit quantum flux, and the
relevant regime is $-\pi <\alpha <\pi$. Further on we set the lattice
spacing $a_0=1$, as well as $h=k_B=1$.

In the following we restrict our study of the model Eq.~(\ref{model})
to the case of a single hole doped into a magnetic insulator. The idea
is that results for a single hole (spin polaron) remain relevant for
the regime of finite, but low, hole concentration $c_h \ll 1$. Here a
semiconductor-like picture implies, so that most measurable
quantities, assuming the independence of spin polarons, should simply
scale with $c_h$. E.g., the diamagnetic susceptibility should behave
as $\chi\propto c_h$.

The ground state of the spin polaron at $B=0$ has been studied
extensively both by analytical and numerical approaches, and can be
considered as one of few rather settled problems within the theory of
correlated systems. Still, here at least two substantially different
regimes have to be distinguished.

At finite $J>0$ (as relevant for cuprates with $J/t\sim0.3$) the
ground state of a hole in an antiferromagnetic (AFM) spin background
has the property of a quasiparticle (QP) with $S=1/2$ and a well
defined dispersion $\varepsilon_0(\vc{k})$. Consistent results have
been obtained for $\varepsilon_0(\vc{k})$ using the self-consistent
Born approximation (SCBA) \cite{schm,mart}, perturbation expansion
\cite{prel}, numerical approaches including both the exact
diagonalization of small systems and the quantum Monte Carlo method
\cite{dago}. Calculations reproduce a minimum at
$\vc{k}^*=(\pm\pi/2,\pm\pi/2)$, which is very anisotropic, i.e.\
$\mu=m_{\perp}/m_{\parallel}\sim5$ for $J/t\sim0.3$. This indicates a
very weak dispersion along the AFM zone boundary, connecting
$\vc{k}=\vc{k}^*$ with $\vc{k}=\vc{k}^{**}=(\pi,0)$,
$(0,\pi)$. Studying small systems \cite{dago} the latter dispersion is
not easy to reproduce correctly. E.g., on a frequently studied system
of $4\times4$ sites, states with $\vc{k}^*$ and $\vc{k}^{**}$ are
degenerate, so pronounced finite size effects are expected. Since a
small $B$ just probes the effective mass of the QP, it is not
surprising that results obtained on small lattices are not reliable or
can be even misleading \cite{bera}. On the other hand, ARPES
measurements on undoped cuprates \cite{well} show a more isotropic
minimum around $\vc{k}^*$. The explanation seems to be beyond the
simple $t$-$J$ model, and the additional effect is attributed to the
next-nearest-neighbor hopping ($t'$) term
\cite{hybe}.

The behavior at $J=0$ is quite different. As shown by Nagaoka
\cite{naga}, the ground state is ferromagnetic (FM) with $S=S_{\rm max}$
and momentum $\vc{k}=0$, where the QP is a simple hole in the filled
band of polarized electrons with an unrenormalized band mass.
Nevertheless, close to this simple QP branch there is a large density
of complicated excited states, where the hole motion is predominantly
incoherent \cite{brin}. Therefore it is expected that even moderate
temperature $T>0$ should have a considerable effect.

The paper is organized as follows. Section II is devoted to the study
of a single hole at $J=0$ and arbitrary $B$. Results are obtained via
the high-$T$ expansion and the Lanczos diagonalization technique for
small systems at $T=0$ and for $T>0$ as well. In Sec.~III we consider
the AFM case with $J>0$. Here the analysis of small systems at finite
$B$ is employed together with the study of the ground state using the
cumulant expansion in $t/J$, which has proven to be very informative
for $B=0$ \cite{prel}. In the last section, Sec.~IV, our results are
summarized and a brief discussion on the magnitude of the
susceptibility and its relation to the Hall constant is presented.

\section{$J=0$ case}

\subsection{High-$T$ expansion}

To study a single hole, as described by the model Eq.(\ref{model})
with $J=0$ and $B>0$, we first use the standard high-$T$ expansion
(HTE). Its application is in this case simple, since the only
expansion parameter is $t/T$, while $B$ remains arbitrary. The free
energy
\begin{equation}
F=-T\,{\rm ln}\, Z=-T\,{\rm ln}\,{\rm Tr}\,
e^{-\tilde{\beta}\tilde{H}},\label{free}
\end{equation}
is within the high temperature expansion expressed in terms of moments
$\mu_n$ and cumulants $\lambda_n$
\begin{eqnarray}
{\rm ln}\, Z&=&{\rm ln}\,{\rm Tr}\,1+{\rm
ln}\,\left[1+\sum_{n=1}^\infty{\frac{\tilde{\beta}^n}{n!}\mu_n}\right]
\nonumber \\
&=& {\rm ln}\,{\rm Tr}\,1+
\sum_{n=1}^\infty{\frac{\tilde{\beta}^n}{n!}\lambda_n},\label{lnz}
\end{eqnarray}
where $\tilde \beta=t/T$, $\tilde H=H/t$ and
\begin{equation}
\mu_n=(-1)^n {\rm Tr}\, \tilde H^n/{\rm Tr}\,1.\label{mu}
\end{equation}

Moments $\mu_n$ can be expressed as a sum over ${n\choose n/2}^2$
closed graphs (paths). Counting different spin configurations for
$B=0$ which remain unchanged \cite{brin} after the performed path,
each graph contributes a weight $2^{f-r+1}$. Here $f$ is the number of
cycles in the spin permutation resulting from a hole traversing the
graph, and $r$ is the number of different sites in the graph. For $B>0$
the only change comes from the contribution of the enclosed magnetic
flux, so that the weight becomes
\begin{equation}
w_n=2^{f-r+1}\,e^{im\alpha},
\end{equation}
where $m$ is the area of the graph in units of $a_0^2$. 

Here it is helpful to choose a $45^\circ$ rotated coordinate system so
that a 2D graph decouples into a direct product of two 1D graphs. In
this way it is straightforward to generate nonequivalent graphs
numerically. We were able to evaluate $\mu_n$ and $\lambda_n$ up to
the order $n=18$. In Table \ref{tab1} lowest cumulants ($n\le 6$)
$\lambda_n=\sum_m\lambda_{mn}\cos m\alpha$ are presented for
illustration, while higher cumulants are available upon request.

\begin{table}
\caption{Cumulants $\lambda_n=\sum_m{\lambda_{nm}\cos m\alpha}$.}
\begin{tabular}{cccccccc}
$\lambda_{nm}$&0&$1$&2&3&4&5&6\\\hline
\r{0}&1&&&&&&\\
\r{2}&4&&&&&&\\
\r{4}&-20&2&&&&&\\
\r{6}&472&-48&$\f{3}{2}$&&&&\\
\r{8}&-24518&5992&-198&$\f{3}{2}$&$\f{1}{4}$&&\\
\r{10}&2207234&$-\f{2703635}{4}$&$\f{65195}{2}$&$-\f{1065}{2}$&
$-\f{1155}{32}$&$\f{5}{8}$&$\f{5}{32}$
\end{tabular}
\label{tab1}
\end{table}

From Eqs.(\ref{free},\ref{lnz},\ref{mu}) the series for the orbital
susceptibility (per one hole) can be generated,
\begin{equation}
\chi=\mu_0{\partial^2 F \over \partial
B^2}{\bigg\vert}_{B=0},\label{defchi}
\end{equation}
\begin{equation}
\frac{\chi}{\chi_0}=-\frac{1}{12}\tilde{\beta}^{3}+
\frac{13}{120}\tilde{\beta}^{5}-
\frac{2087}{16128}\tilde{\beta}^{7}+\frac{8161}{53760}\tilde
{\beta}^{9}-\cdots,\label{chi}
\end{equation}
where $\chi_0=\mu_0e^2a_0^4t/\hbar^2$.

There is no unique procedure for the extrapolation of the power series
Eqs.(\ref{lnz},\ref{chi}) to low $T$. For the present problem the most
reasonable approach seems to be via the density of states
$\rho(\varepsilon)$ and their moments \cite{brin},
\begin{eqnarray}
Z&=&\int{\rho(\varepsilon)\,e^{-\beta\varepsilon}{\rm
d}\varepsilon},\nonumber \\
\mu_n&=&\int{\varepsilon^n\rho(\varepsilon){\rm
d}\varepsilon}.\label{dens}
\end{eqnarray}

The density of states can be expanded in terms of Legendre
polynomials,
\begin{equation}
\rho(\varepsilon)=\sum_{\ell=0}^\infty{B_\ell{\rm
P}_\ell(\varepsilon)},\qquad
\mu_n=\sum_{\ell=0}^n{C_{n\ell}B_\ell},\label{leg}
\end{equation}
with coefficients
\begin{equation}
C_{n\ell}={\Gamma(\f{n}{2}+\f{1}{2})\Gamma(\f{n}{2}+1)\over
2\Gamma(\f{n}{2}+\f{\ell}{2}+\f{3}{2})
\Gamma(\f{n}{2}-\f{\ell}{2}+1)},\label{coef}
\end{equation}
for $n>\ell$ and even $\ell+m$, while $C_{n\ell}=0$ otherwise.

The density of states is used to extrapolate both $F$ and $\chi$ to low 
$T$. After solving the linear equations
(\ref{leg}) for $B_\ell$, we can calculate the susceptibility
\begin{equation}
\frac{\chi}{\chi_0}=\frac{1}{Z\tilde\beta}
\left[\frac{\partial^2\!Z}{\partial\alpha^2}-
\frac{1}{Z}\left(\frac{\partial Z}{\partial\alpha}\right)^2\right],
\label{chit}
\end{equation}
where the same set of equations as in (\ref{dens}-\ref{coef}) holds
for derivatives with respect to $\alpha$ as well.

\subsection{Small system diagonalization}

In analysis of the $t$-$J$ model the Lanczos technique for the
exact diagonalization of small systems has been already extensively
employed \cite{dago}, predominantly in the investigation of the ground
state and their static and dynamic properties. Recently a method
combining the Lanczos procedure and the random sampling has been
introduced \cite{jakl1} which allows calculation of finite-temperature
properties in small correlated systems. The method has been used in
the study of various response function within the $t$-$J$ model at
$T>0$ \cite{jakl1}. The application is particularly simple for static
quantities, which can be expressed as expectation values of conserved
quantities \cite{jakl2}. The calculational effort is comparable to the
ground state evaluation. In particular the average energy $\langle
E\rangle$,
\begin{eqnarray}
\langle E\rangle&\approx&\frac{N_{st}}{KZ}\sum_{n=1}^{K}
\sum_{m=0}^{M-1}|\langle n|\psi_m^n\rangle|^2E_m^n
{\rm e}^{-\beta E_m^n},\nonumber\\
Z&\approx&\frac{N_{st}}{K}\sum_{n=1}^{K}
\sum_{m=0}^{M-1}|\langle n|\psi_m^n\rangle|^2{\rm e}^{-\beta E_m^n},
\label{lan}
\end{eqnarray}
should be evaluated in this way, where $|\psi_m^n\rangle, E_m^n$ are
respectively approximate eigenfunctions and energies obtained by the
diagonalization within the orthonormal set, generated from the initial
functions $|n\rangle$ in $M$ Lanczos steps. The $K$ initial functions
$|n\rangle$ are chosen at random, while $N_{st}$ is the dimension of
the complete basis. Note that it is enough to choose $M,K
\ll N_{st}$. For more detailed explanations we refer to
Refs.\cite{jakl1,jakl2}.

The introduction of finite $B>0$ in the model, Eq.(\ref{model}),
reduces the translational symmetry and thus for a given system size
increases the required minimal basis set. We are at present able to
consider the problem of a single mobile hole in canonical 2D systems
with $N = 16,18,20$ sites \cite{oitm}, and periodic boundary
conditions (p.b.c.).

It is nontrivial to incorporate phases due to a homogeneous $B$, being
at the same time compatible with p.b.c. It is well known \cite{frad}
that this can be accomplished only for quantized magnetic fields $B=m
B_0$, where $B_0=\phi_0/N$ is the smallest field corresponding to the
unit quantum flux per system. To incorporate such $B$ in small systems
(tilted squares), the following procedure is used: a) phases
$\theta_{ij}$ for all hops inside squares are left as given within the
particular Landau gauge, b) phases attributed to hops across the
square boundaries are subject to the condition that the magnetic flux
in each plaquette remain the same, i.e.\ $B'\equiv B~\rm{mod}(\phi_0)$
(up to the addition of an unit flux per plaquette). These boundary
requirements lead to a set of linear equations which have solutions
only for $B=mB_0$. Only for an untilted square lattice with $N=L\times
L$ sites the phases can be expressed in a simple form as
\begin{eqnarray}
\theta_{(i_x,i_y)(i_x,i_y+1)}&=& B i_x , \nonumber\\
\theta_{(i_x,i_y)(i_x+1,i_y)}&=& 0,\qquad\quad i_x<L,\nonumber\\
\theta_{(L,i_y)(1,i_y)}&=& B i_y L.
\label{bound}
\end{eqnarray}

\subsection{Results}

Let us first discuss the polaron internal energy
$\langle\varepsilon(\alpha)\rangle$ as obtained from the high-$T$
expansion
\begin{equation}
\langle\varepsilon\rangle=-{\partial({\rm ln}\,Z)\over\partial\beta}=
-t\sum _{n=0}^\infty{{\tilde\beta}^n \over n!}\lambda_{n+1}.
\label{energ}
\end{equation}
From cumulants $\lambda_{nm}$ in Table I we see that the $\alpha$
dependence of $\langle\varepsilon\rangle$ first enters within the
order $\tilde\beta^3$. Such a term originates from a hole hopping
around a loop, contributing to ${\rm Tr}\,\tilde H^4$ when all spins
are equally polarized, in analogy to the processes contributing to the
Hall constant \cite{brin1}. As a result, $\alpha$ dependence of
$\langle\varepsilon\rangle$ is vanishing fast ($\propto(t/T)^3$) for
$T>t$.

\begin{figure}[t]
\begin{center}
\iffigure
\epsfig{file=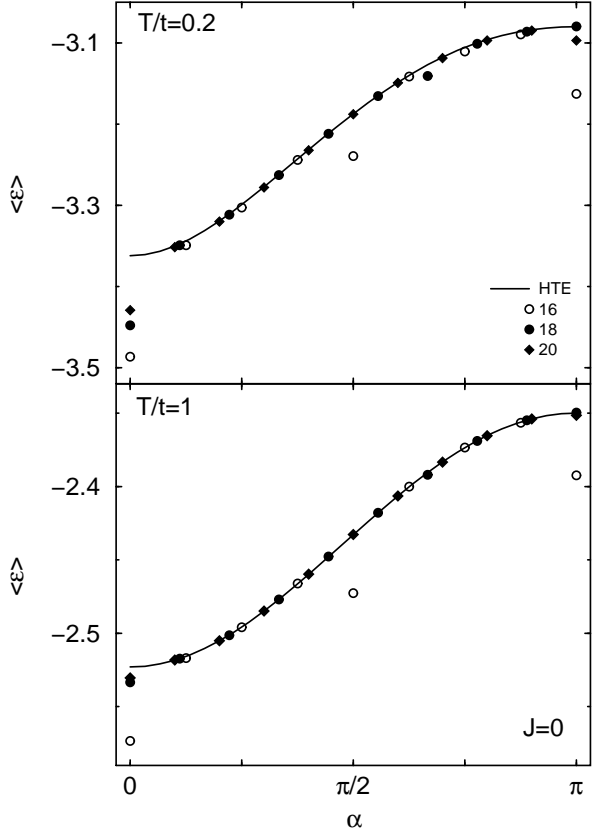,width=8cm}
\fi
\end{center}
\caption{Polaron energy $\langle\varepsilon\rangle$ (in units of $t$)
for $J=0$ at two different temperatures, as a function of
dimensionless magnetic field $\alpha$. Full line represents high-$T$
expansion results, while open and full dots correspond to results on
systems with $N=16,18$ and $20$ sites, respectively.}
\label{fig1}
\end{figure}

In Fig.~\ref{fig1} we present $\langle\varepsilon(\alpha)\rangle$ for
lower $T$, i.e.\ $T=t$ and $T=0.2~t$. The results of the high-$T$
extrapolation and finite size calculations performed for $N=16,18$ and
20 are compared. For most points the agreement between both methods is
quite satisfactory. On the other hand, there are clearly visible
anomalies at $\alpha = n \pi/2$ for $N=16$ and less pronounced at
$\alpha=0$ for $N=18, 20$. It is straightforward to explain these
discrepancies within the high-$T$ expansion. In small systems there
are some additional processes due to p.b.c., which lead to changes of
$\langle\varepsilon(\alpha)\rangle$ relative to an infinite system.
E.g., in the $N=4\times4$ system contributions from graphs in lowest
field-dependent order $\tilde{\beta}^3$, representing four consecutive
hops in the $x$- or $y$-direction cancel, except for $\alpha= n
\pi/2$, changing the cumulant
\begin{equation}
\Delta\lambda_4=1,
\end{equation}
as seen in Fig.~\ref{fig1}.

It should be observed that for $N=16$ this correction to $\langle
\varepsilon\rangle$ is within the leading order of $\tilde\beta^3$,
while analogous corrections in larger systems, e.g.~for $N=18$, emerge
only in higher orders. This confirms that on small systems the
calculation of $B$-induced diamagnetic currents is more delicate than
the evaluation of most of the static polaron properties. Nevertheless,
at $J=0$ finite size effects are rather well under control, at least
in comparison to the AFM case $J>0$ presented in Section III.

\begin{figure}[t]
\begin{center}
\iffigure
\epsfig{file=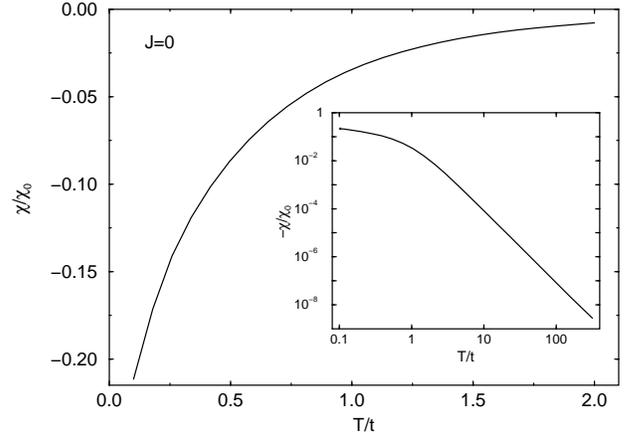,height=8cm,angle=-90}
\fi
\end{center}
\caption{Susceptibility $\chi$ vs.~$T/t$, obtained by the high-$T$
expansion.}
\label{fig2}
\end{figure}

Consistency of high-$T$ expansion and small system ($N>16$) results
allows reliable extrapolation of the susceptibility $\chi$ to quite low
$T\sim 0.1~t$, using the procedure via the density of states
$\rho(\varepsilon)$, Eqs.(\ref{dens}-\ref{chit}). The result is
presented in Fig.~\ref{fig2}, and as expected $\chi$ is
diamagnetic. While for $T\gg t$ the susceptibility $\chi$ is
proportional to $\beta^3$, the variation is less steep for
$0.1~t<T<t$, where the variation is closer to
$\chi\propto\beta^{\eta}$ with $\eta<1$.

It is quite delicate to approach $T=0$ within the $J=0$ model. The
ground state for a single hole is within the sector of maximal total
spin $S=S_{\rm max}$ \cite{naga}, however excited states are numerous
and close in energy, so that the transition between the regime of an
incoherent hole propagation and the regime of the large FM polaron
appears to happen at surprisingly low $T^*/t \sim 0.1$. This is
consistent with the well established fact that the FM-polarized ground
state is very sensitive to any change of parameters. The behavior of a
single hole is simple only strictly at $T=0$. It is expected that the
QP at $B=0$ behaves according to the Nagaoka theorem, i.e.\ as a free
hole with energy $\varepsilon=-4~t$ in a filled band of spinless
fermions. Looking only in the sector $S=S_{\rm max}$, a finite field
($B>0$) should increase the ground state energy according to the
cyclotron frequency, i.e.\ $\delta
\varepsilon\sim eB/m^*=Bt$.

\begin{figure}[t]
\begin{center}
\iffigure
\epsfig{file=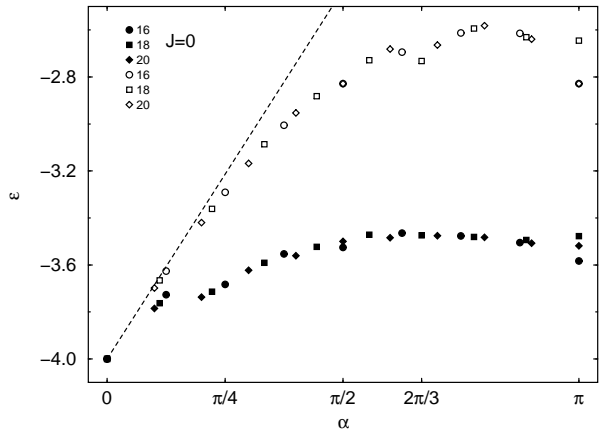,height=8cm,angle=-90}
\fi
\end{center}
\caption{Ground state energy $\varepsilon$ (in units of $t$)
vs.~$\alpha$, obtained via the diagonalization of small systems of
different sizes and $J=0$. Open dots correspond to the lowest-energy
state within the sector $S=S_{\rm max}$.}
\label{fig3}
\end{figure}

In Fig.~\ref{fig3} we show both the absolute g.s.\ energy
$\varepsilon$ and the lowest energy in the sector $S=S_{\rm max}$ as
calculated within small systems with $N=16,18$ and 20. For the Nagaoka
sector $S=S_{\rm max}$ the behavior is for $\alpha < \pi/4$ clearly of
the cyclotron type, while for higher $\alpha$ there are some visible
commensurability anomalies identical to the study of spinless fermions
in a magnetic field \cite{kohm}. On the other hand, deviations of the
absolute ground state from the naive result are much more
pronounced. First, even the smallest $B=B_0$ leads to the instability
of the $S=S_{\rm max}$ state, and the actual spin of the g.s.\ is
$S<S_{\rm max}$. Nevertheless, $\delta\varepsilon$ remains quite close
to the cyclotron value. For higher $B>B_0$ the ground state saturates
quite abruptly to lowest spin $S=1/2$ and g.s.\ energy $\varepsilon$
is much lower than in the $S=S_{\rm max}$ case. Note, however, that
even an approximate validity of the simple cyclotron-frequency
argument, $\delta\varepsilon\propto|B|$, implies a divergent
susceptibility $\chi(T$$\to$$0)\to\infty$, as found also from the
high-$T$ expansion.

\section{$J>0$: Antiferromagnetic spin background}

$J>0$ introduces quite a different QP behavior. At $T=0$ the spin
background corresponds to an AFM with long range order. The ground
state of the spin polaron is well understood, and corresponds within
the $t$-$J$ model to $S=1/2$ and $\vc{k}=\vc{k}^*$, with a weak
dispersion along the AFM zone boundary.

At least for weak magnetic fields it is expected that the QP
description of the hole is still valid, resulting in a cyclotron
motion with the linear-in-field dependence of the QP energy. Such
behavior, is questioned by recent small-system diagonalisation results
\cite{bera}. It should be reminded that results obtained from
the exact diagonalisation can be misleading due to finite size
effects, if the cyclotron radius is comparable to the linear size of
the system. With p.b.c.~even more pronounced effects compared to those
already discussed for $J=0$ in Sec.~II can be expected. For the AFM
case they arise from larger ground state degeneracy, and from the
near-degeneracy of the QP dispersion along the AFM zone boundary.

\subsection{Cumulant expansion method}

At $T$=$0$ we can consider a QP moving in an ordered AFM by performing
the expansion starting in the limit $t/J \ll 1$. Here the standard
cumulant-expansion (CE) procedure for the ground-state energy, as
first considered by C.~Bloch \cite{bloc}, is followed.

The implementation of the method in the zero-field case with a single
hole in the $t$-$J$ model has been already given in some detail
elsewhere \cite{prel}. There the diagonal part of the Heisenberg spin
interaction in Eq.~(1) has been taken as the unperturbed Hamiltonian
$H_0$ and the ground-state energy $\varepsilon_0(\vc{k})$, expressed
relative to the undoped AFM ground state, has been obtained as a
double power series in $u=t/J$ and $\gamma = J_\perp /J$
\begin{equation}
\varepsilon_0(\vc{k})=J\sum_{n,m}a_{n,m}(\vc{k})u^n({\gamma\over
2})^m\,,
\end{equation}
where the $\vc{k}$-dependence of the expansion coefficients merely
reflects the underlying translational invariance of the problem.
 
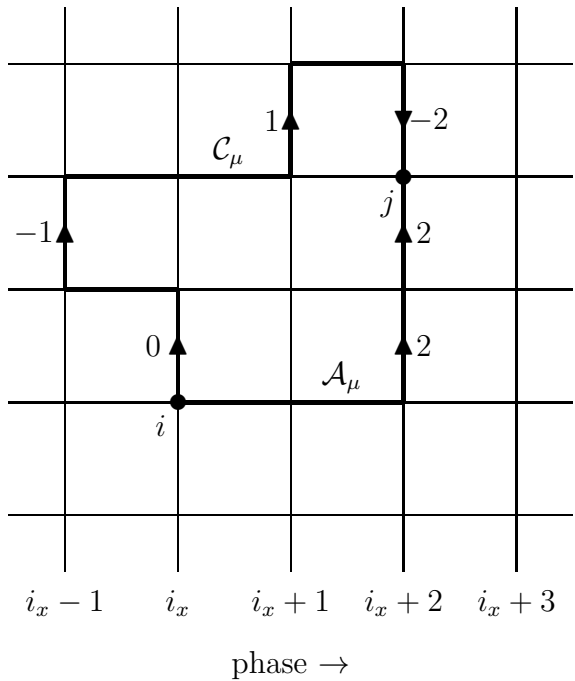
\begin{figure}[t]
\def\cir{1.3}
\def\mysize#1{{\large #1}}
\def\btr{\blacktriangleright}
\def\btl{\blacktriangleleft}
\def\btd{\blacktriangledown}
\def\btu{\blacktriangle}
\begin{center}
\setlength{\unitlength}{1.5mm}
\begin{picture}(60,60)(0,-5)
\linethickness{0.2mm}
  \multiput(10,5)(10,0){5}{\line(0,1){50}}
  \multiput(5,10)(0,10){5}{\line(1,0){50}}
\linethickness{0.5mm}
\Huge
  \put(40,20){\line(0,1){30}}
  \put(20,20){\line(1,0){20}}
  \put(20,20){\line(0,1){10}}
  \put(20,30){\line(-1,0){10}}
  \put(10,30){\line(0,1){10}}
  \put(10,40){\line(1,0){20}}
  \put(30,40){\line(0,1){10}}
  \put(30,50){\line(1,0){10}}
  \put(20,20){\circle*{\cir}}
  \put(40,40){\circle*{\cir}}
  %\put(25,20){\makebox(0,0){\mysize{$\btr$}}}
  \put(40,25){\makebox(0,0){\mysize{$\btu$}}}
  \put(40,35){\makebox(0,0){\mysize{$\btu$}}}
  \put(20,25){\makebox(0,0){\mysize{$\btu$}}}
  \put(40,45){\makebox(0,0){\mysize{$\btd$}}}
  \put(10,35){\makebox(0,0){\mysize{$\btu$}}}
  \put(30,45){\makebox(0,0){\mysize{$\btu$}}}
  \put(18,17){\makebox(0,0)[lb]{\mysize{$i$}}}
  \put(38,37){\makebox(0,0)[lb]{\mysize{$j$}}}
  \put(9,35){\makebox(0,0)[r]{\mysize{$-1$}}}
  \put(18.5,25){\makebox(0,0)[r]{\mysize{$0$}}}
  \put(29,45){\makebox(0,0)[r]{\mysize{$1$}}}
  \put(42.5,25){\makebox(0,0)[r]{\mysize{$2$}}}
  \put(42.5,35){\makebox(0,0)[r]{\mysize{$2$}}}
  \put(44,45){\makebox(0,0)[r]{\mysize{$-2$}}}
  \put(24.5,43.5){\makebox(0,0)[t]{\mysize{$\cal{C}_\mu$}}}
  \put(34.5,23.5){\makebox(0,0)[t]{\mysize{$\cal{A}_\mu$}}}
  \put(10,1.5){\makebox(0,0)[b]{\mysize{$i_x-1$}}}
  \put(20,1.5){\makebox(0,0)[b]{\mysize{$i_x$}}}
  \put(30,1.5){\makebox(0,0)[b]{\mysize{$i_x+1$}}}
  \put(40,1.5){\makebox(0,0)[b]{\mysize{$i_x+2$}}}
  \put(50,1.5){\makebox(0,0)[b]{\mysize{$i_x+3$}}}
  \put(30,-4){\makebox(0,0)[b]{\mysize{phase $\to$}}}
\end{picture}
\end{center}
\caption{The phase increments, defined on the (oriented) links between
two neighboring lattice points, in the Landau gauge $\vc{A}=B(0,x,0)$
along two different paths ${\cal C}_\mu$ and ${\cal A}_\mu$
connecting points $i$ and $j$.}
\label{fig4}
\end{figure}

A finite magnetic field breaks the translational invariance of the
system, so the above result, obtained from the nondegenerate theory is
not valid. Indeed, the unperturbed ground state becomes highly
degenerate, allowing for any hole position in an otherwise perfectly
N\'eel-ordered spin state, thus the degenerate perturbation theory
must be used. The version of the secular determinant as given in
\cite{mess} is used, which up to the fourth order in perturbation $H'$
reads
\begin{eqnarray}
(E&-&E_0)\delta_{ij}\nonumber\\&=&<i|H'{Q_0\over \eta_0}H'|j>+
<i|H'{Q_0\over\eta_0}H'{Q_0\over\eta_0}H'{Q_0\over \eta_0}H'|j>\nonumber\\
&-&\sum_\ell<i|H'{Q_0\over\eta_0}H'|\ell><\ell |H'{Q_0\over 
\eta_0^2}H'|j>\nonumber\\
&+&\cdots
\label{dpex}
\end{eqnarray}
The (bra)kets denote the set of degenerate states -- positions of the
hole -- with the unperturbed energy $E_0$, whereas $P_0$ and $Q_0$ are
the projectors onto this subset of states and its complement,
respectively. $H'$ is taken to be the sum of the hopping part and the
transverse spin part of $H$ in (1), while $\eta_0$ stands for the
energy denominator $E_0-H_0$. A contribution to the matrix element
$M(i,j)=<i|\cdots |j>$ resulting from a particular process $\cal
O_\mu$ involving products of operators as in (\ref{dpex}) is
calculated from the associated graph \cite{prel,bloc} in any order of
perturbation expansion. However, as the hole moves from some initial
position $i$ to a final but equivalent position $j$ under the action
of a particular string of operators $\cal O_\mu$ along some path, it
acquires a definite phase factor $\theta_{ij}$, depending on the path
itself. Two such paths are depicted in Fig.~\ref{fig4}.

In the chosen Landau gauge the phase associated with any link in the
$y$-direction is given by the $x$-coordinate of that link. Thus, the
total phase $\theta_{ij}$ acquired by the hole along the path $\cal
C$, is
\begin{equation}
\theta_{ij}({\cal C})=\int_{\cal C}\vc{A}\cdot {\rm d}\vc{s}
= i_x(j_y-i_y)\alpha + \phi_{ij}({\cal C}).
\end{equation}
Here $\phi_{ij}({\cal C})$ is the phase relative to the initial point
at $i$. Care should be taken of the proper orientation in which the
link is traversed. Thus, referring to Fig.~\ref{fig4}, phase along
${\cal C}_\mu$ is $\phi= -2\alpha$, whereas along the path ${\cal
A}_\mu$ $\phi= 4\alpha$.

We have generated all the paths in order $u^n\gamma^m$ with $(n,m)$ up
to $(2,4)$, $(4,4)$, $(6,3)$ and $(8,2)$. Each contribution to the
perturbation series in order $(n,m)$ is then given by the magnitude
$\omega(i,j)$ of the matrix element $\langle i|{\cal O}_\mu^{(r)}|
j\rangle$ and the phase $\theta_\mu(i,j)$, where ${\cal O}_\mu^{(r)}$
refers to a definite product of operators of order $r=n+m$ along the
path from $i\to j$ \cite{prel}.

The secular equation (\ref{dpex}) becomes a difference equation for the
on-site amplitudes $f_i$ on a rectangular grid with $N=L_x\times L_y$
sites
\begin{eqnarray}
\varepsilon f_i&=&\sum_{j} M(i,j) f_j,\nonumber\\
M(i,j)&=&\sum_{n,m}u^n\gamma^m \sum_{{\cal C}_\mu}\omega_\mu(i,j )
e^{\imath\theta_\mu(i,j)},\label{sr2d}\nonumber\\
\end{eqnarray}
where $i$ and $j$ run only over one sublattice, $M(i,j)$ is a sum of
contributions along different paths ${\cal C}_\mu$, and the energy
$\varepsilon\equiv\varepsilon(\alpha)$ is again measured with respect
to the g.s.\ of the undoped AFM state. Referring again to
Fig.~\ref{fig4} the paths ${\cal C}_\mu$ and ${\cal A}_\mu$ would then
first appear in order $u^8\gamma^4$ and $u^4\gamma^2$, respectively.

In the chosen gauge Eq.~(\ref{theta}), the system is translationally
invariant along the $y$-direction. Thus, the ansatz
$f_i=g_{i_x}\exp(\imath k_y i_y)$ reduces the above equation to a
difference equation in one dimension, where $\ell=i_x$,
\begin{eqnarray}
\varepsilon g_\ell=&&\sum_{\ell'} {\widetilde 
M}(\ell,\ell')g_{\ell'},\qquad 1\le \ell,\ell'\le
L_x,\nonumber\\
{\widetilde M}(\ell,\ell')={\sum_\tau}'e^{\imath\tau 
Q}&&\sum_{n,m}u^{n}\gamma^m
\sum_{{\cal 
C}_\mu}\omega_\mu(\ell'-\ell,\tau)e^{\imath\phi_\mu(\ell'-\ell,\tau)}
\nonumber\\ Q=&&k_y+\alpha \ell,\label{eige}
\end{eqnarray}
and the summation over $\tau$ is restricted to run over values for which
$\ell'-\ell+\tau$ is even, whereas $k_y\in
\lbrack0,2\pi\rbrack$. Note also that $\omega$ and $\phi$ do not
depend on the initial point $\ell$, but only on the path $\cal
C$. Taking $\alpha=2\pi p/L_x$, one can impose the p.b.c.~also in the
$x$ direction.
 
\begin{figure}[t]
\begin{center}
\iffigure
\epsfig{file=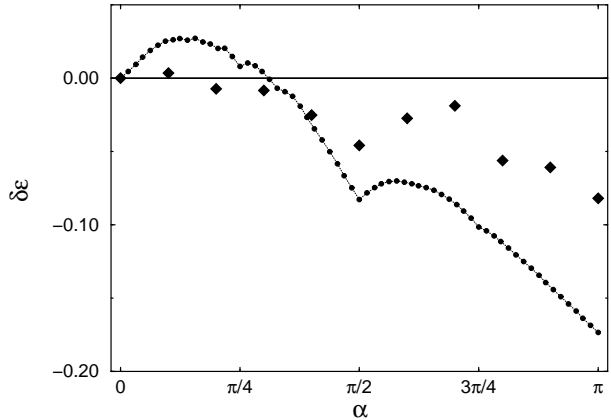,height=8cm,angle=-90}
\fi
\end{center}
\caption{Ground state energy difference
$\delta\varepsilon(\alpha)=\varepsilon(\alpha)-\varepsilon(0)$
(in units of $t$), as obtained for $t/J=0.5$ and $\gamma=1$ from
the cumulant expansion method. For comparison small-system data for
$N=20$
are also included ($\blacklozenge$)}
\label{fig5}
\end{figure}

The eigenvalue problem of Eq.(\ref{eige}) is solved numerically for
$L_x\gg\xi$, where $\xi=13$ is the smallest linear size of the region
visited by the hole, within the order of perturbation series here
considered. In Fig.~\ref{fig5} we plot the g.s.\ energy $\varepsilon$
as a function of $\alpha$ at $J/t=2$ and the isotropic exchange
$\gamma=1$, evaluated for $L_x=128$. A linear-in-field dependence of
$\delta\varepsilon(\alpha)=\varepsilon(\alpha)-\varepsilon(0)$ in
Fig.~\ref{fig5} is evident for small $\alpha$, implying that the hole
may still be described as a QP exhibiting cyclotron motion. However,
after the initial rise an almost monotonic decrease is observed and
the minimum of $\varepsilon(\alpha)$ is achieved for $\alpha=\pi$.
The figure includes also the respective data from exact
diagonalization for $N=20$. Although both sets of data do not agree in
detail, which can be partially attributed to the perturbational
character of the CE results, the overall behavior is remarkably
similar, including the initial rise in $\delta\varepsilon(\alpha)$ for
$N=20$ and the cusp-like behavior close to $\alpha=\pi/2$. This holds
to a lesser degree for other commensurability points, e.g.\ for
$\alpha=\pi/4$ and $3\pi/4$.

The experimentally relevant value of $J/t$ is $\sim 0.3$.
Technically, it is possible to perform a Pad{\' e}-like extrapolation
of $\varepsilon(\alpha)$ to larger values of $u$, in analogy to the
$B=0$ case \cite{prel}. Still we do not attempt to perform such an
extrapolation here. Relying on previous experience
\cite{prel} that no crossover in the QP behavior sets in down to $J/t
\ll 1$ (where Nagaoka regime takes over) we believe that the qualitative
conclusions on QP behavior for finite $B$ remain valid in the
physically relevant regime $J<t$ as well.
 
\subsection{Small-system results}

Using the numerical finite temperature technique described in Sec.~IIB
and applied to the $J=0$ case the results for $J>0$ are also
obtained. In the following we choose $J/t=0.4$, which is close to the
situation in cuprates, and consider the contribution of the single
hole energy $\varepsilon=E(N_h$=$1)- E(N_h$=$0)$ to the internal
energy at $T=0$ and $T>0$ ($\langle\varepsilon\rangle$). First note
that for $T\gg {\rm max}(t,J)$, the leading order of the high-$T$
expansion is independent of $J$ and in this limit results for $J=0$
and $J>0$ match.

\begin{figure}[t]
\begin{center}
\iffigure
\epsfig{file=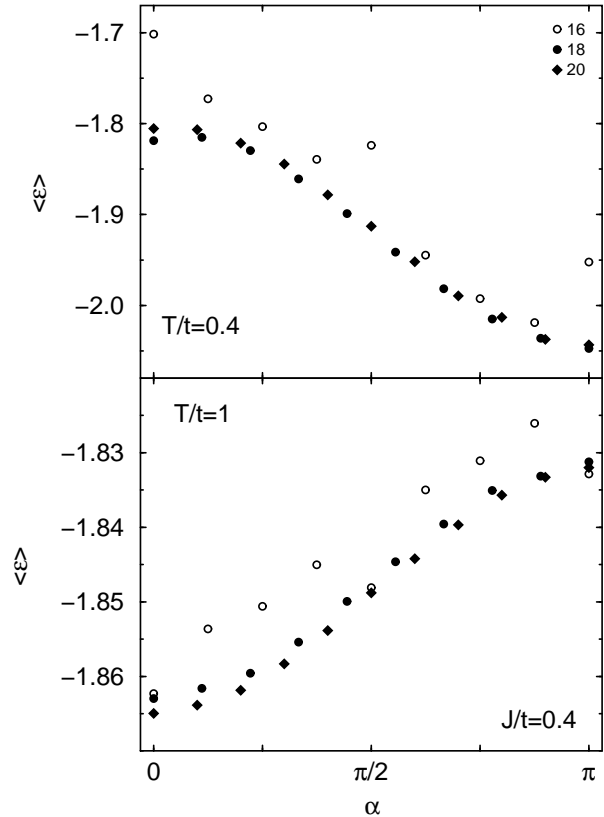,width=8cm}
\fi
\end{center}
\caption{Single hole energy $\langle\varepsilon\rangle$ vs.~$\alpha$ for
$J/t=0.4$ at two different $T/t$, as calculated on a system of $N=16,
18, 20$ sites.}
\label{fig7}
\end{figure}

In Fig.~\ref{fig7} the intermediate temperature hole energies
$\langle\varepsilon(\alpha)\rangle$ are evaluated at two different $T$
within different system sizes. Comparing results with those for $J=0$
in Fig.~\ref{fig1}, several conclusions can be reached. At higher
$T/t=1$ energies $\langle\varepsilon(\alpha)\rangle$ for $J=0$ and
$J/t=0.4$ are qualitatively similar. As a function of $\alpha$ both
cases correspond approximately to the simple $\cos
\alpha$ variation. It is however evident that finite $J$ considerably
reduces (by a factor $\sim 4$) the total energy span.

Relative to the $J=0$ case it is also clear that finite size effects
become more pronounced when approaching the low-$T$ regime. This is
related to the near-degeneracy of the quasiparticle dispersion
$\varepsilon_0({\vc k})$ at the bottom of the QP band, which is poorly
reproduced in small systems. Nevertheless, numerical results establish
quite consistently the crossover at $T\sim T^*\propto J$.

For $T<T^*$ energy $\langle\varepsilon(\alpha)\rangle$ shows an
overall opposite, i.e.\ a paramagnetic-like variation with $\alpha$,
as found in \cite{bera} and reproduced above within the CE with
respect to $t/J$. The results in this temperature region are more size
dependent. At $T/t=0.2$ in Fig.~\ref{fig7} we can find a maximum for
$\alpha>0$ only in systems with $N=18,20$ sites, while the $N=16$ case
shows a different behavior. The deviation within the latter system is
due to additional symmetry of the $4\times4$ system (hypercube).

\begin{figure}[t]
\begin{center}
\iffigure
\epsfig{file=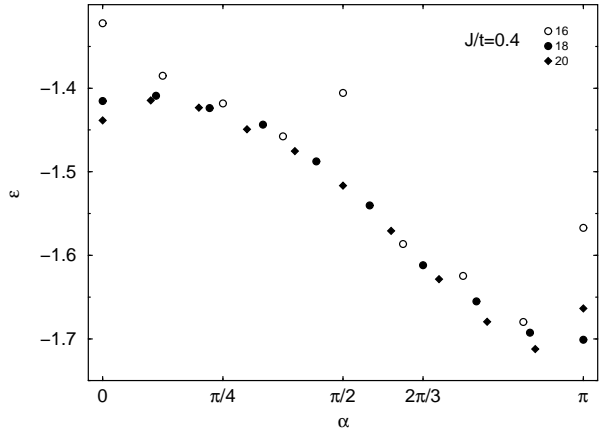,height=8cm,angle=-90}
\fi
\end{center}
\caption{Ground state energy $\varepsilon(\alpha)$ for
$J/t=0.4$, as calculated on systems of
$N=16, 18, 20$ sites.}
\label{fig8}
\end{figure}

Quite similar results are obtained for the ground state energy
$\varepsilon$, shown in Fig.~\ref{fig8}. Again, $N=18$ and $N=20$
systems yield very similar $\varepsilon(\alpha)$, while $N=16$ results
deviate especially for `commensurate' values $\alpha=0,\pi/2, \pi$.
The qualitative trend of $\varepsilon(\alpha)$ agrees well with the CE
results on Fig.~5. The similarity is in the existence of (rather
shallow) maximum at low $\alpha$ for largest $N$, indicating a QP
behavior with a cyclotron motion in weak $B$ and also a pronounced
reduction of $\varepsilon$ for $\alpha >\pi/2$. The only difference is in
the absence of commensurability dips in Fig.~8, which we attribute to
the very particular system shapes used in the calculation, and to the
almost degenerate dispersion along the magnetic Brillouin zone
boundary.
 
\begin{figure}[b]
\begin{center}
\iffigure
\epsfig{file=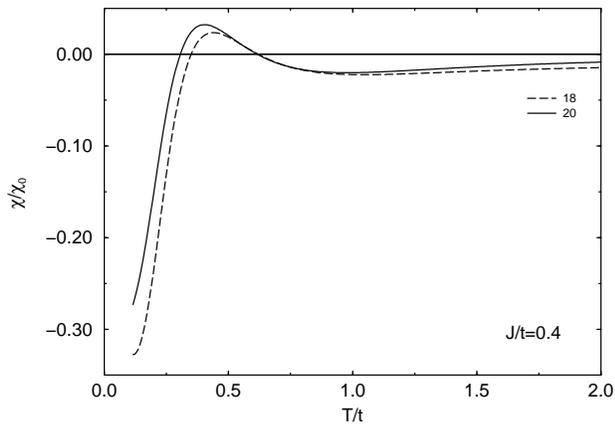,height=8cm,angle=-90}
\fi
\end{center}
\caption{Susceptibility $\chi$ vs.~$T/t$ for $J/t=0.4$, as extracted
from small-system data $N=18,20$.}
\label{fig9}
\end{figure}

Finally, let us present in Fig.~\ref{fig9} results for the orbital
susceptibility $\chi(T)$, Eq.(\ref{defchi}). For $J=0$ the calculation
is performed via high-$T$ expansion as already discussed in Sec.~2.
Since we rely only on discrete values of $\alpha$, the analysis of
finite-system data for $J/t=0.4$ is on the other hand less reliable.

Moreover, the variation of $F(\alpha)$ is quite subtle at smallest
$\alpha_{\rm min}$, as evidenced also from $\varepsilon$ in
Fig.~\ref{fig8}. In Fig.~\ref{fig9} the susceptibility $\chi$ is
presented as obtained from Eq.(\ref{defchi}), with parabolic fit for
$F(\alpha)$ using only $\alpha=0$ and $\alpha=\alpha_{\rm min}$. This
procedure could be questionable for $T<T^*$, but qualitative behavior
is still quite instructive. Relative to the $J=0$ case, the
diamagnetism is suppressed by $J>0$ at higher $T\gg J$. In an
intermediate regime $T\sim J$, susceptibility $\chi$ appears even to
change sign, i.e.\ becomes paramagnetic. Only at low $T<T^*$ a
pronounced strongly $T$-dependent diamagnetic response is again
observed, consistent with the QP cyclotron motion at $T=0$. It is an
interesting observation that an exact solution of the problem on a
single plaquette (at $T>0$) subjected to an effective staggered AFM
field, reproduces qualitative features of Fig.~\ref{fig9}.

\section{Conclusions}

Our study shows that the calculation of the effects of finite magnetic
field, coupled to orbital motion of electrons, becomes delicate in
models of correlated electrons, e.g.\ within the $t$-$J$ model
considered in this work. Results for magnetic observables such as the
diamagnetic susceptibility appear to be strongly influenced by
finite-size effects, which are hard to overcome in available system
sizes. Both for $J=0$ and $J>0$ some of deviations appear at
`commensurate' values of $\alpha$ within the given system geometry,
and are particularly large for $N=16$. We have shown that it is
possible to understand such finite-size effects within the high-$T$
expansion or within the $t/J$ perturbation expansion as a contribution
of additional graphs due to p.b.c. Still it is impossible to
eliminate them systematically in most interesting physical
regimes. These effects lead to a nonmonotonous variation of
observables, e.g.\ $\langle\varepsilon(\alpha)\rangle$, which in turn
leads to an enhanced uncertainty in $\chi(T)$.

The $J=0$ case seems both easier to study and to understand. High-$T$
expansion and small systems show a continuous transition from the
high-$T$ regime of incoherent hopping to the Nagaoka state at $T=0$,
with a monotonous increase of the diamagnetic $\chi$. At $T>0$ where
the HTE is reliable the variation of the energy with field
$\langle\varepsilon(\alpha)\rangle$ is quite close to a simple
$\cos\alpha$ form. Nevertheless the asymptotic behavior at low $T$ is
not simple to establish, since the nature of low lying states (above
the Nagaoka state) is complicated.

The behavior of the AFM $J>0$ polaron is more involved. While at $T\gg
J$ the exchange scale $J$ is not important and results qualitatively
follow those for $J=0$, new physics emerges for $T\lesssim J$. A
nearly flat $\langle\varepsilon(\alpha)\rangle$ at intermediate regime
$T\sim J$ is quite remarkable, and leads to a vanishing diamagnetic
$\chi$ (or even change of its sign). It seems, that here $J>0$
diminishes and even destroys emerging coherence of QP. Only at lower
$T<J$ the coherence is established and the known dynamical picture of
a coherent AFM polaron is dominating the behavior in lowest fields
$\alpha$. Reliable results are however difficult to obtain even for
$T=0$, since in small systems their variation with $\alpha$ is very
sensitive to the system shape, boundary conditions etc.\ due to the
very anisotropic and degenerate QP dispersion.

The CE results are very instructive, but it is not straightforward to
make them quantitative for $J<t$. For $J>t$, where the CE series
converges quite rapidly, there is a clear evidence in weak fields that
the QP is exhibiting cyclotron motion. The commensurability effects
are quite pronounced and agree with finite-cluster data for large
$J/t>1$, see Fig.~\ref{fig5}. On the other hand, these effects are not
evident in small-system results for $J/t<1$. This non-agreement is
attributed in part to specific lattice shapes and p.b.c. The flatness
of the dispersion seems to be the main reason for the structureless
character of small-system data in the region of small $\alpha$ for
$J/t<1$.

Here we would like to point to the close similarity with the
challenging theoretical problem of the Hall effect, which is difficult
to approach even for very low doping, e.g.\ for a single hole
\cite{prel1}. Similarly to the orbital susceptibility the Hall effect
emerges due to coupling to orbital currents. The Hall constant is
given by $R_H=\sigma_{xy}/B\sigma_{xx}^2$, where the off-diagonal
conductivity $\sigma_{xy}$ can be related to the orbital
susceptibility $\chi$ as
\cite{rojo}
\begin{equation}
\sigma_{xy}=B\frac{\partial\chi}{\partial c_h}
\frac{\partial c_h}{\partial \mu},
\end{equation}
where $\mu$ is the chemical potential. In our case of very low hole
doping, i.e.\ in the semiconductor-like regime, the susceptibility
$\chi$ scales linearly with $c_h$; moreover $\partial
c_h/\partial\mu=-\beta c_h$, so that $\sigma_{xy}\propto\chi$. The
high-$T$ expansion of the Hall constant $R_H^*(T)$ (the high-frequency
value) is analogous to that of $\chi(T)$ \cite{brin1}. On the other
hand, crossing the scale $T\sim J$ remains the challenge, whereby it
seems that at this intermediate $T$ the hole-like $R^*_H$ is even
reduced with respect to its high-$T$ value \cite{brin1,assa}.
Experiments \cite{ong} indicate, that $R_H$ recovers for $T<T^*$,
varying strongly with $T$ and approaches the well known quasiclassical
result for $T\to 0$ \cite{prel1}. Note that our results for $\chi(T)$,
Fig.~\ref{fig9}, indicate just on such behavior.

Let us finally comment on the magnitude of the diamagnetic
susceptibility. Since we are evaluating the case of a single hole, at
low doping $c_h\ll 1$ the observable diamagnetic contribution to the
susceptibility (per unit cell) should be $\chi=\zeta c_h\chi_0$, where
$\zeta$ is dimensionless value, presented in Figs.~\ref{fig2}
and~\ref{fig9}. It is convenient to compare these values to the spin
susceptibility (per unit cell) of the planar undoped AFM for $T<T^*$,
where $\chi_s\sim4.0\mu_0\mu_B/J$ \cite{jakl2,sing}. Setting
$m_t=\hbar^2/2ta_0^2$ the ratio can be expressed as
\begin{equation}
\frac{\chi}{\chi_s}=K\zeta c_h\frac{J}{t}\left(\frac{m_e}{m_t}\right)^2,
\label{rat1}
\end{equation}
where $K\sim 2.8$ is a numerical constant. Taking the standard values
for cuprates $t=0.4$ eV, $J/t=0.3$ and $a_0=0.38$ nm, the above
relation reduces to $\chi/\chi_s\sim1.9\,c_h\,\zeta$.

Another relation can be obtained for the Pauli susceptibility of a
half-filled band of free tight-binding electrons, where a constant
(average) density of states is assumed for simplicity. This gives a
similar relation
\begin{equation}
\frac{\chi}{\chi_P}=K'\zeta c_h\left(\frac{m_e}{m_t}\right)^2
\sim9.3\,c_h\,\zeta\,,
\label{rat2}
\end{equation}
with $K'\sim 4.0$. To estimate the actual value of $\chi/\chi_s$ in
Eq.~(27) we take $c_h\sim 0.15$, e.g.\ as in the 'optimal' doping
regime, and $\zeta\sim-0.1$ in the region $T<T^*$. This gives
$\chi/\chi_s\sim-0.03$, which is of the same order of magnitude as the
experimentally measured value \cite{wals}. Note, however, that below
the crossover temperature $T^*$ $\zeta$ becomes strongly temperature
dependent, as opposed to the usual $T$-independent Landau-type
diamagnetism in Fermi liquids. Since it is difficult to distinguish
different contributions to the actual susceptibility in experiments,
it remains to be seen whether such $T$-dependent $\chi$ really appears
in cuprates and analogous systems.

\acknowledgements

One of the authors (P.P.) wishes to thank X.~Zotos for helpful
suggestions concerning the introduction of a magnetic field in small
systems.

This work was supported by Ministry of Science and Technology of
Slovenia under Project No.\ J1-6166-0106/97.

\vspace{5mm}
* E-mail: \verb|darko.veberic@ijs.si|


\begin{references}
\bibitem{prip} Note that within the CE approach the g.s.\ is AFM-like to
any finite order of the perturbation series. The hole can thus visit
sites of one sublattice only.
\end{references}
\end{document}